\documentclass[superscriptaddress,aps,prd,showpacs,preprint,longbibliography]{revtex4-1}

\usepackage{lingmacros}
\usepackage{tree-dvips}
\usepackage{graphicx}
\usepackage{amsmath}
\usepackage{amssymb}
\usepackage{latexsym}
\usepackage{color}
\usepackage{fullpage}
\usepackage{commath}
\usepackage{cancel}
\usepackage{ulem}
\usepackage{natbib}
\usepackage{dcolumn}
\usepackage[latin1]{inputenc}
\usepackage{graphicx, psfrag}
\usepackage{float}
\usepackage{tensor}
\usepackage{amsfonts}
\usepackage{hyperref}
\usepackage{subfigure}
\usepackage{pgfplots}
\usepackage{epstopdf}
\usepackage{booktabs}
\usepackage[11pt]{moresize}

\begin{document}

\title{Noether Charge, Thermodynamics and Phase Transition of a Black Hole in the Schwarzschild-Anti-de Sitter-Beltrami spacetime}

\author{T. Angsachon} \email{tossang@tu.ac.th} 
\affiliation{Department of Physics, Faculty of Science and Technology, Thammasat University 99 Moo 18 Paholyothin Road, Klong Nueng, Klong Luang, Pathumthani 12121, Thailand}

\author{K. Ruenearom} \email{f.krerkpon@gmail.com}
\affiliation{Department of Physics, Faculty of Science and Technology, Thammasat University 99 Moo 18 Paholyothin Road, Klong Nueng, Klong Luang, Pathumthani 12121, Thailand}

\begin{abstract}
In this work, we investigate thermodynamic properties and Hawking-Page phase transition of a black hole in the Schwarzschil-Anti-de Sitter-Beltrami (SAdSB) spacetime. We discuss the Beltrami or inertial coordinates of the Anti-de Sitter(AdS) spacetime. Transformation between the non-inertial and inertial coordinates of the AdS spacetime is formulated to construct the solution of spherical gravitating mass and other physical quantities. The Killing vector is determined to calculate the event horizon radius of this black hole. The entropy and the temperature of SAdSB black hole are determined by the Noether charge method and it is shown that the temperature is bounded by the Anti-de Sitter radius. Similarly, the Smarr relation and first law of black hole thermodynamics for the SAdSB spacetime have been formulated. The Gibbs free energy and heat capacity of this black hole are calculated and we consider the phase transition between small and large black holes. The first-order phase transition between the thermal AdS spacetime and large black hole phase is also investigated and the Hawking-Page temperature is computed and compared with the case of the Schwarzschild-Anti-de Sitter(SAdS) black hole.
\end{abstract}

\maketitle{}

\newpage
\section{Introduction}

\qquad From classical general relativity, nothing can escape from black hole's event horizon. However considering the quantum fluctuations around the event horizon, black holes can emit particles. Therefore, they can have temperature and entropy, and can be considered as thermal objects \cite{Bekenstein:1973ur, PhysRevLett.26.1344, Hawking:1975vcx}. The thermodynamic properties of black holes relates to their mechanical properties by the laws of black hole mechanics \cite{Bardeen:1973gs, PhysRevD.13.191} which suggests that the temperature and entropy of a black hole can be associated to its surface gravity and horizon area, respectively. As a result, the laws of the black hole thermodynamics can be formulated by the methods of the Noether charge with the variation of gravitational field action
\cite{Wald:1993nt, PhysRevD.50.846, PhysRevD.49.6587, Frolov:1998wf,PhysRevD.74.044007}.

A spherically symmetric black hole solution with positive(negative) cosmological constant was discovered by F. Kottler \cite{1918AnP...361..401K} and it is called the Schwarzschild-(Anti)-de Sitter (S(A)dS) spacetime. The thermodynamics of the Schwarzschild black hole in Anti-de Sitter spacetime (SAdS) was first investigated in \cite{Hawking:1982dh}. Interestingly, there are two phases of the SAdS black holes, i.e., small and large black holes with negative and positive heat capacity, respectively. Moreover, the transition between the two stable phases, namely the thermal radiation and large black hole phases, at a certain temperature emerges.  It is known as the first order Hawking-Page phase transition. Recently, a number of studies are devoted to the duality between the AdS spacetime geometry and the conformal field theory, which is known as the AdS/CFT correspondence \cite{Maldacena:1997re,Witten:1998qj,Witten:1998zw,Aharony:1999ti}. This duality has a huge influence to the community in studying the non-perturbative quantum chromodynamics (QCD) via the thermodynamics of the SAdS black hole. Especially, the Hawking-Page phase transition can be used as the holographic dual to the confinement/deconfinement phase transition  \cite{Maldacena:1997re,Witten:1998qj,Witten:1998zw,Aharony:1999ti}.

%{\color{red} Talk about Robert Mann's work and road to our result}

According to \cite{Kastor:2009wy,Dolan:2012jh,Kubiznak:2012wp,Kubiznak:2016qmn}, the cosmological constant $(\Lambda)$ can be interpreted as pressure $P=-\Lambda/{8\pi}$ such that the work term can be presented and completes the first law of black hole mechanics. It is proposed that the SAdS black hole has a positive pressure induced by the negative value of $\Lambda$. Then the thermodynamic volume of the SAdS black hole can be determined as the derivative of the mass with respect to the pressure. In this way, the mass of the SAdS black hole could be interpreted as the chemical enthalpy. This approach can be called the extended phase space approach of the black hole in the AdS spacetime \cite{Wang:2020hjw}. Moreover, in the AdS/CFT context, the value of $\Lambda$ corresponds to the number of coincident branes $N$ in the bulk spacetime. The variation of the pressure leads to the variation the number of colors $N$. The equation of state depending on the field degrees of freedom $N$ and the thermal phase transition in conformal field theory has been determined and demonstrated in \cite{Johnson:2014yja,Dolan:2014cja,Dolan:2016jjc}. % (describe gauge gravity)} 
Recently, a novel thermal equilibrium can occur by considering the R\'enyi statistics of the asymptotically flat black holes \cite{Czinner:2015eyk, Czinner:2017tjq, Promsiri:2020jga, PhysRevD.104.064004, Promsiri:2022qin} and asymptotically dS spacetime \cite{Tannukij:2020njz, Samart:2020klx, Nakarachinda:2021jxd}. As suggested in \cite{Promsiri:2020jga, PhysRevD.104.064004}, the non-extensivity parameter can be interpreted as a thermodynamic pressure and its conjugate as a thermodynamic volume. This novel framework is known as the R\'enyi extended phase space approach.    

It is crucial to determine the coordinates which represent the inertial reference frame in de-Sitter and Anti-de Sitter spacetimes. This coordinate system is known as the Beltrami coordinates \cite{Guo:2003qm, Yan:2005wf, Guo:2007gz,Manida:2011qg,2015dsis.book.....Y}, and also called the (Anti)-de Sitter-Beltrami ((A)dSB) spacetime. Because this spacetime has a group isometry, we can construct the generator of symmetry and conserved quantities of this spacetime with the help of the Noether charge method in the same manner as in Minkowski spacetime \cite{Angsachon:2013lzk}. Because of the inertial effect, the geodesic line in the (A)dSB spacetime is a straight line similar to that of in the Minkowski spacetime. By the group contraction method, the generators of symmetry in the AdSB spacetime can be transformed to the generators of symmetry in the Minkowski spacetime \cite{Manida:2011qg}. Furthermore, the Beltrami coordinates of the (A)dS spacetime have the boundary in radial coordinate $r$, which is different from the static coordinates of (A)dS spacetime that the radial coordinate $r$ does not have boundary \cite{Guo:2003qm, Yan:2005wf, Guo:2007gz,Manida:2011qg,2015dsis.book.....Y}. Moreover, the Beltrami time coordinate does not contain a periodic imaginary character. It means that a zero-temperature state should be determined in (A)dSB spacetime \cite{Guo:2004xg}. The Schwarzschild (Anti)-de Sitter-Beltrami spacetime was first studied in \cite{Angsachon:2013bx,Sun:2013vfa}, which shows that the Schwarzschild-Anti-de Sitter metric in non-inertial or Beltrami coordinates can be constructed via the spherical coordinate transformation from the non-inertial or static Schwarzschild-Anti-de Sitter metric.
Thermodynamics of the Schwarzschild-de Sitter-Beltrami (SdSB)  black hole was investigated in \cite{Liu:2016urf} and shown that black holes with positive curvature appear only in an unstable state in Beltrami coordinates. 

Hence in this paper, we investigate thermodynamic properties of a black hole in Schwarzschild-Anti-de Sitter-Beltrami (SAdSB) spacetime. The thermodynamic quantities of this spacetime can be constructed from the existence of the generator representing the timelike symmetry. We suggest that the entropy of the SAdSB black hole can be determined by the method of the Noether charge \cite{Wald:1993nt,PhysRevD.50.846, PhysRevD.49.6587, Frolov:1998wf}, and the first law of black hole thermodynamics can be formulated in a similar way.
Remarkably, even in the inertial Beltrami frame, there are unstable phase (small black hole) and stable phases (large black hole and thermal radiation). Furthermore, we expect that the SAdSB black hole can be considered as thermodynamic system in a box or a cavity due to the existence of the radial boundary in the AdSB spacetime.

This paper is organized as follows. In Section 2, the Beltrami coordinates in Anti-de Sitter spacetime is introduced and reviewed. We describe the geodesic equation and determine the time translation generator in AdSB spacetime. In Section 3, we determine the transformations between spherical non-inertial or static and spherical inertial (Beltrami) coordinates. In Section 4, we construct the metric of the Schwarzschild-AdSB spacetime by using the coordinate transformation between the AdS and AdSB spacetime. We consider the event horizon of the SAdSB black hole determined by a norm of the time-translation Killing vector of the AdSB spacetime. Section 5 is devoted to determine the Noether charges from the time-translation Killing vector which is used to calculate the obtained entropy of the SAdSB black hole. We show that the entropy obeys to the Bekenstein-Hawking area law. In section 6, we demonstrate that the black hole mass can be considered as the enthalpy while the Hawking temperature is related to the surface gravity. The heat capacity and the Gibbs free energy is calculated to demonstrate all possible phases of this black hole.  We discuss Hawking-Page phase transition and free-energy landscape. Throughout this work, we set the constant values $c = \hbar = G_4 = k_B = 1$.
%Interestingly to research the mass solution of the AdSB space
%- History of BH thermodynamics and AdS/CFT correspondence
%- History about Beltrami
%- Motivation/Question in ours works
%- Ours result
%- Paper organize

\section{Anti-de Sitter-Beltrami Spacetime}

It is known the $AdS_4$ spacetime is represented as the embedded hyperboloid 
\begin{eqnarray}
X^2_{-1} + X^2_{0} - X^2_{1} - X^2_{2} - X^2_{3}= R^2,
\end{eqnarray}
where $R$ is the parameter describing a radius of the AdS spacetime.\\
So the line element of the $AdS_4$ is represented by the five-dimensional ambient Minkowski spacetime as \cite{Aharony:1999ti}
\begin{equation}\label{em}
ds^2 = dX_AdX^A = dX_{-1}^2+dX_0^2-dX_1^2-dX_2^2-dX_3^2,
\end{equation}
where the index $A$ runs from $-1,0,1,2,3$. 

Now we introduce the Beltrami coordinates by taking the projection from the center of the five-dimensional embedded hyperboloid on the hyperplane on which $X_{-1} = R$ with the parametrization as
\begin{eqnarray}\label{tr}
x_\mu = R\frac{X_\mu}{X_{-1}},
\end{eqnarray}
where $x_\mu$ are the Beltrami coordinates \cite{Guo:2003qm,Yan:2005wf,Guo:2007gz,2015dsis.book.....Y,Manida:2011qg,Angsachon:2013lzk} and the index $\mu$ runs from $0,1,2,3$. By substituting the coordinate transformation \eqref{tr} to the metric (\ref{em}), we obtain the line element as
\begin{eqnarray}\label{AdSB}
ds^2 = \frac{\eta_{\mu \nu}dx^\mu dx^\nu}{h^2}-\frac{(\eta_{\mu \nu}x^\mu dx^\nu)^2}{R^2 h^4},
\end{eqnarray}
where $h^2 = 1+\frac{\eta_{\mu\nu} x^\mu x^\nu}{R^2}$ and $\eta_{\mu\nu}$ is the metric tensor of the Minkowski spacetime. Then the spacetime  described by the line element \eqref{AdSB} is called as the anti-de Sitter-Beltrami(AdSB) spacetime \cite{Manida:2011qg,Angsachon:2013lzk}.

So the Ricci scalar of this spacetime is constant and equal to $\mathcal{R} = -12/R^2$ that is the value of the Ricci scalar in the AdS spacetime. This means that the AdSB is also a spacetime with a constant curvature.
This metric tensor satisfies the solution of the Einstein-Hilbert action with the negative cosmological constant
\begin{equation}\label{act}
I = -\frac{1}{16\pi}\int d^4 x \sqrt{-g}(\mathcal{R}-2 \Lambda),
\end{equation}
where $\Lambda = -\frac{3}{R^2}$ is the relation between the cosmological constant and the AdS radius.

The geodesic equation for this spacetime can be written as
\begin{eqnarray}
\frac{d^2 x^\mu}{ds^2}-\frac{2}{R^2 h^2}\frac{dx^\mu}{ds}(x^\nu\frac{dx^\nu}{ds}) = 0.
\end{eqnarray}
Interestingly, the solution of this equation is given by
\begin{eqnarray}
x^i = v^i t+c^i,
\end{eqnarray}
where $v^i = \frac{dx^i}{dt}$ and $c^i$ is an arbitrary constant. This implies that the trajectory of a massive particle in the AdSB spacetime is a straight line in the similar way as a free particle in Minkowski spacetime \cite{Guo:2003qm,Yan:2005wf,Guo:2007gz,2015dsis.book.....Y,Angsachon:2013lzk} although the curvature of this spacetime is not zero. It means that the Beltrami coordinates \eqref{tr} can be considered as the inertial coordinates in AdS spacetime. Therefore, the inertial reference frame can be constructed in the constant curvature spacetime such as the AdSB spacetime. 

We see that the AdSB spacetime has a boundary at $h^2 = 0$, so the boundary radius is determined as $r_b = \sqrt{(\eta_{ij}x^ix^j)}|_b = \sqrt{R^2+t^2}$. It means that the radial domain of this spacetime realized under the interval $r~\in~[0,r_b]$. It is different from the AdS spacetime in which the radial domain has unbounded value $r~\in~[0,\infty)$ \cite{Aharony:1999ti}.

%The corresponding isometry group (the relativity group of dS4) is SO(1, 4), i.e. the Lorentz group of the ambient spacetime M(1,4) which is generated by the following ten Killing vector fields. 
It is known that the symmetry group of the embedded $AdS_4$ spacetime is $SO(3,2)$ which is realized by the generators of symmetry or the Killing vectors expressed as \cite{Angsachon:2013lzk}
\begin{equation}\label{gen}
L_{AB} = X_A\frac{\partial}{\partial X^B}-X_B\frac{\partial}{\partial X^A}.
\end{equation}
We are interested to consider the generator of transformation between the timelike embedded coordinates $X_{-1}$ and $X_0$ which represents time translation in four-dimensional Beltrami coordinate.
So the generator of time translation in the Beltrami coordinates is given as \cite{Manida:2011qg,Angsachon:2013lzk} 
\begin{equation}\label{tgen}
H = \left(1+\frac{t^2}{R^2}\right)\frac{\partial}{\partial t}+\frac{tx_i}{R^2}\frac{\partial}{\partial x^i},
\end{equation}
which leads to the conserved energy of a massive particle in the AdSB spacetime \cite{Angsachon:2013lzk}. The generator \eqref{tgen} is analogous to time translation generator ($\frac{\partial}{\partial t})$ in the Minkowski spacetime which also expresses the conserved energy of a massive particle. Similarly, the other conserved quantity such as linear and angular momentum had been also constructed from the generator symmetry \eqref{gen} \cite{Angsachon:2013lzk}. It means that the Beltrami coordinates of the AdS spacetime not only shows the existence of the initial frame reference, but also there are the physical conserved quantity in this spacetime in the same way as in the flat Minkowski spacetime. 

Moreover, the Beltrami time generator \eqref{tgen} is contracted to the Minkowski time generator under the limit $R\longrightarrow \infty$. Likewise, by the method of deformation of the Galilean group algebra the time translation  ($\frac{\partial}{\partial t})$ also converts to the time translation generator \eqref{tgen} \cite{Manida:2011qg}. Thus, only the Beltrami coordinates of the Anti-de Sitter spacetime can be contracted by the group structure to the Minkowski spacetime.  
 Furthermore, we will show that by the coordinate transformation between the AdS and AdSB spacetimes which will be discussed in the section 3. And later in section 4, we will show that the generator \eqref{tgen} is the timelike Killing vector in the AdSB spacetime.
 
% Discuss symmetry, generator of symmetry Killing vector Inertial frame of observer in AdS space. BH from this observer. Conserved quantity which can define the BH thermo in AdSB.

\section{Spherical Coordinate transformation between the anti-de Sitter-Beltrami and the anti-de Sitter spacetimes}

In this section, we derive the coordinates transformation between the non-inertial coordinates system in AdS spacetime $x^\mu = (t_n, r_n, \theta_n, \phi_n)$ and the inertial coordinates system in AdSB spacetime $x^\mu = (t, r, \theta, \phi)$. We start to consider the metric of the AdS spacetime \cite{Hawking:1982dh, Witten:1998qj,Witten:1998zw,Aharony:1999ti}
\begin{equation}\label{AdS coordinates}
ds^2 = \left(1+\frac{r_n^2}{R^2} \right)dt^2 - \left(1+\frac{r_n^2}{R^2} \right)^{-1}dr_n^2 - r_n^2 (d\theta_n^2 + \sin^2\theta_n d\phi_n^2). 
\end{equation}
The metric of the AdSB spacetime in \eqref{AdSB} can be written in the form of spherical coordinates as
\begin{equation}\label{AdsB coordinates}
ds^2 = \frac{1}{h^2}\left(1-\frac{t^2}{R^2h^2}\right)dt^2+\frac{2rt}{r^2h^4}drdt-\frac{1}{h^2}\left(1+\frac{r^2}{R^2h^2}\right)dr^2-\frac{r^2}{h^2}(d\theta^2 + \sin^2\theta d\phi^2), 
\end{equation}
where 
\begin{eqnarray}
h = \sqrt{1+\frac{t^2-r^2}{R^2}}. 
\end{eqnarray}
We assume that the time and spatial coordinates in the non-inertial system should be written in the inertial coordinates system as $t_n = t_n(t)$, $r_n = r_n(r,t)$ and $\theta_n = \theta_n(\theta)$, $\phi_n = \phi_n(\phi)$. Since the metric tensor transformation rule is
\begin{equation}\label{tran}
g_{\mu\nu} = \frac{\partial x^\rho_n}{\partial x^\mu}\frac{\partial x^\sigma_n}{\partial x^\nu}g_{\mu\nu}^n,
\end{equation}
where $g_{\mu\nu}$ and $g_{\mu\nu}^n$ represents the metric tensor of the AdSB spacetime and the metric tensor of the AdS spacetime respectively, and hence we put the metric \eqref{AdS coordinates} and \eqref{AdsB coordinates} to the transformation \eqref{tran}. We obtain the system of the differential equations of the AdSB and  AdS coordinates representation as 
\begin{eqnarray}\label{adstr}
  \frac{1}{h^2}\left(1-\frac{t^2}{R^2h^2}\right)   &=&  \left(\frac{\partial t_n}{\partial t}\right)^2\left(1+\frac{r_n^2}{R^2}\right)-\left(\frac{\partial r_n}{\partial t}\right)^2\left(\frac{1}{1+\displaystyle\frac{r_n^2}{R^2}}\right), \\
    \frac{1}{h^2}\left(1+\frac{r^2}{R^2h^2}\right)   &=&   \left(\frac{\partial r_n}{\partial t}\right)^2\left(\frac{1}{1+\displaystyle\frac{r_n^2}{R^2}}\right), \\
    \frac{r^2}{h^2} &=& \left(\frac{\partial \theta_n}{\partial \theta}\right)^2 r_n^2, \\
     \label{adstr1}
    \frac{r^2}{h^2}\sin\theta &=& \left(\frac{\partial \phi_n}{\partial \phi}\right)^2 r_n^2 \sin\theta_n.
\end{eqnarray}
The non-trivial solution of \eqref{adstr}-\eqref{adstr1} are
\begin{eqnarray}\label{bel1}
t_n &=& L \arctan \left(\frac{t}{R}\right), \\
\label{bel2}
r_n &=& \frac{r}{\sqrt{1+\displaystyle\frac{t^2 - r^2}{R^2}}}, \\
\label{bel3}
\theta_n &=& \theta, \\
\label{bel4}
\phi_n &=& \phi. \label{bel5}
\end{eqnarray}
This is the coordinate transformation between the AdS and AdSB spacetimes.
It means that any the physical metric tensor in AdSB spacetime can be constructed from the static coordinates of the AdS spacetime through this coordinate transformation and we will describe in detail in next section. 

\section{The Schwarzschild-Anti-de Sitter-Beltrami spacetime}

In this section, we investigate the Schwarzschild-AdSB spacetime and formulate the line element of a spherical symmetric gravitating mass $M$ in the inertial frame of the spacetime with the negative constant curvature. The original SAdS metric can be describe as \cite{1918AnP...361..401K, Hawking:1982dh,Witten:1998qj, Witten:1998zw, Aharony:1999ti}
\begin{eqnarray}\label{SAdS}
 ds^2 = \left( 1+\frac{r_n^2}{R^2}-\frac{2 M}{r_n} \right)dt_n^2-\left( 1+\frac{r_n^2}{R^2}-\frac{2 M}{r_n} \right)^{-1}dr_n^2 - r_n^2(d\theta_n^2 + \sin^2\theta_n d\phi_n^2). 
\end{eqnarray}
In order to obtain the SAdSB metric, we use the transformation rules in \eqref{bel1}, \eqref{bel2}, \eqref{bel3} and \eqref{bel4}, the metric in \eqref{SAdS} becomes
\begin{eqnarray}\label{SAdSB}
ds^2=\left(\frac{1}{h_0^4}f(r,t)-\frac{r^2 t^2}{R^4h^6f(r,t)}\right)dt^2+2\frac{h_0^2tr}{R^2h^6f(r,t)}dtdr
-\frac{h_0^4}{h^6f(r,t)}dr^2 - \frac{r^2}{h^2}d\Omega^2,
\end{eqnarray}
where
\begin{eqnarray}
d\Omega^2 &=& d\theta^2 + \sin^2\theta d\phi^2, \\
h &=& \sqrt{1+\frac{t^2-r^2}{R^2}},\\
h_0 &=& \sqrt{1+\frac{t^2}{R^2}},  \\
f(r,t) &=& 1+ \frac{r^2}{R^2h^2} - \frac{2M h}{r}.
\end{eqnarray}
This metric describes the spherically symmetric solution with a negative cosmological constant in an inertial reference frame \cite{Angsachon:2013bx,Sun:2013vfa}.

It is crucial to explore the black hole solution in the Beltrami coordinate representation. Since the event horizon of a black hole is a Killing horizon which is a region where a Killing vector field is normal, the norm of Killing vector field goes to zero on the event horizon surface of black hole \cite{Frolov:1998wf}. It is known that the Killing vector of the AdS space is the time translation generator, namely $\xi_n = \frac{\partial}{\partial t_n}$ which can be transformed into the Beltrami coordinates via the transformation \eqref{bel1}-\eqref{bel2} as
\begin{equation}\label{kill}
\xi = \frac{\partial x^\mu}{\partial t_n}\frac{\partial}{\partial x^\mu} = h_0^2\frac{\partial}{\partial t} + \frac{rt}{R^2}\frac{\partial}{\partial r},
\end{equation}
or in the components form
\begin{equation}\label{kill1}
\xi^\mu = \left( h_0^2, \frac{rt}{R^2}, 0, 0 \right).
\end{equation}
Interestingly, the given Killing vector \eqref{kill} corresponds to the time-translation generator of the AdSB spacetime in spherical coordinates \eqref{tgen}. Moreover, the AdSB Killing vector \eqref{kill1} obeys to the Killing equation
\begin{equation}
\nabla^\mu \xi^\nu+\nabla^\nu \xi^\mu = 2\nabla^{[\mu}\xi^{\nu]} = 0.
\end{equation}

The norm of the Killing vector of the SAdSB metric \eqref{SAdSB} is
\begin{equation}\label{norm}
\xi^2 %= g_{\mu \nu}\xi^\mu \xi^\nu
= f(r,t) = 1+\frac{r^2}{R^2h^2}-\frac{2M h}{r}.
\end{equation}

Furthermore, we define the event horizon radius the SAdSB black hole as $r_+$ and see that from the equation \eqref{norm} the event horizon of this black hole is determined by the function $f(r_+,t) = 0$, because the Killing vector is timelike outside the black hole region $f(r>r_+)>0$, it is null at the event horizon $f(r_+)=0$, and it is spacelike inside the black hole $f(r<r_+)<0$. Thus, to obtain the event horizon radius, we introduce a new variable $x_+=\displaystyle\frac{r_+}{h}$ and substitute it into the horizon function
\begin{equation}\label{ev}
f(x_+)=1+\frac{x_+^2}{R^2}-\frac{2 M}{x_+} = 0.
\end{equation}
So the solution of this equation is expressed as
\begin{equation}\label{9}
x_+=\frac{r_+}{h}=\frac{2}{\sqrt{3}}R \sinh\left(\frac{1}{3}\sinh^{-1}\frac{3\sqrt{3} M}{R} \right),
\end{equation}
and we get the event horizon radius which depends on the gravitating mass $M$ and a time variable as
\begin{equation}\label{23}
r_+= \frac{\displaystyle\frac{2}{\sqrt{3}}R\sinh\left(\frac{1}{3}\sinh^{-1}\left(\frac{3\sqrt{3} M}{R}\right)\right)}
{\sqrt{1+\displaystyle\frac{4}{3}\frac{r_b^2}{R^2}\sinh^2 \left(\frac{1}{3}\sinh^{-1}\left(\displaystyle\frac{3\sqrt{3} M}{R}\right)\right)}},
\end{equation}
where $r_b = \sqrt{R^2+t^2}$ is a bounded radius determined by the AdSB background metric (\ref{SAdSB}).
The bounded radius is time-dependent. In the limit $R \gg t$ the radius is slowly changed and the bounded radius approximately to the AdS-radius $r_b \simeq R$.
The relation between the mass and the event horizon radius can be also received from ($\ref{ev}$) as
\begin{equation}\label{23a}
M = \frac{x_+(R^2+x_+^2)}{2R^2} =
\frac{r_+r_b^2R}{2(r_b^2-r_+^2)^{3/2}}.
\end{equation}
Relatively, we see that the mass value of SAdSB black hole is bounded by this radius. Later we will use the mass to define the Smarr relation and the first law of black hole mechanics(thermodynamics) in the sections IV and V.

\section{Entropy of the Schwarzschild Anti-de Sitter-Beltrami Black Hole as a Noether Charge}

In this section we determine the entropy of the Schwarzschild-Anti-de Sitter-Beltrami black hole. Although the metric of the SAdSB black hole depends on time, there are isometry and generator of symmetry described in section II. We can formulate the entropy area law, the Smarr relation and the first law of black hole mechanics by the method of the Noether charge. %Review about Iyer-Wald method for SAdSB BH
R. Wald and V. Iyer proposed \cite{Wald:1993nt,PhysRevD.50.846} that the Lagrangian of a gravitational field which is invariant under diffeomorphism, can be employed to construct the Noether charge. From the Einstein-Hilbert's action with a negative constant curvature in (\ref{act}), the Noether current density can be thus represented as
\cite{PhysRevD.50.846, PhysRevD.49.6587, Frolov:1998wf, PhysRevD.74.044007}
\begin{equation}\label{noe}
J^\mu = \frac{1}{8\pi}\nabla_\nu Q^{\mu\nu},
\end{equation}
where
\begin{equation}\label{Q}
Q^{\mu\nu} = \frac{1}{8\pi}\nabla^{[\mu}\xi^{\nu]} =  \frac{1}{16\pi}(\triangledown^\mu\xi^\nu-\triangledown^\nu\xi^\mu),
\end{equation}
are the Noether charge elements of a gravitational field in the constant curvature spacetime generated by Killing vector $\xi^\mu$ \cite{PhysRevD.74.044007, Urano:2009xn}.

With the integral over the 2-dimensional surface and volume, the Noether charge for the Einstein's gravitational fields in anti-de Sitter spacetime is shown as \cite{PhysRevD.50.846, PhysRevD.49.6587, Frolov:1998wf, PhysRevD.74.044007,Urano:2009xn}
\begin{equation}\label{noe1}
Q = \int_{\partial\Sigma} Q^{\mu \nu} d\sigma_{\mu\nu},
\end{equation}
where %used other notation for Jacobian
\begin{equation}
d\sigma_{\mu\nu} = \frac{1}{4}\sqrt{-g}\varepsilon_{\mu\nu\theta\phi}\text{det}\left(\frac{\partial(x^\mu,x^\nu)}{\partial(\theta,\phi)}\right)d\theta d\phi,
\end{equation}
is the element of a closed two-dimensional surface of the event horizon and $\partial\Sigma$ is the surface over an event horizon area. %Tell the reader to see appendix about Wald formalism in AdS-BH 

Hence, the non-zero elements of the Noether charge density for the SAdSB metric are
\begin{equation}
Q^{01} = -Q^{10} =\frac{1}{16\pi}(\triangledown^0\xi^1-\triangledown^1\xi^0)  =\frac{h^2}{8\pi R^2}\left( r-M h+\frac{M R^2h h_0^2}{r^2} \right),
\end{equation}
while the non-zero components of the two-dimensional integral element are
\begin{equation}
d\sigma_{01} = -d\sigma_{10} = \frac{r^2}{2h^5}\sin \theta d\theta d\phi.
\end{equation}
As a result, we obtain the conserved quantity of the SAdSB black hole from the integral \eqref{noe1} as 
\begin{equation}\label{noe2}
Q = 2\int_{\partial \Sigma} Q^{01} d\sigma_{01}
= \frac{M}{2}+\frac{x_+^3}{2R^2}.
\end{equation}
Furthermore, the entropy of this black hole is also calculated from the integral of the Noether charge over the event horizon surface\cite{PhysRevD.50.846, PhysRevD.49.6587, Frolov:1998wf, PhysRevD.74.044007} with the help of the surface gravity. In the same manner, the entropy for any black hole can be determined through the Noether charge of the gravitational Lagrangian \cite{PhysRevD.50.846, PhysRevD.49.6587, Frolov:1998wf}. Thus the black hole entropy can be formulated as 
\begin{equation}\label{en}
S_\text{BH} = \frac{2\pi}{\kappa_\text{H}}Q =\frac{2\pi}{\kappa_\text{H}}\int_{\partial\Sigma} Q^{\mu\nu}d\sigma_{\mu\nu},
\end{equation}
where $\kappa_H$ is the surface gravity of the SAdSB black hole on the event horizon.\\ 
With the help of the mass relation \eqref{23a} the surface gravity of the SAdSB metric is evaluated using its definition as
\begin{equation}\label{kh}
\kappa_H = \sqrt{-\frac{1}{2}\nabla^\nu \xi^\mu \nabla_\nu \xi_\mu}  =\frac{1}{2}\frac{df(x)}{dx}\bigg|_{x_+} = \frac{R^2+3x_+^2}{2R^2 x_+}.
\end{equation}

Therefore, the entropy in \eqref{en} of the SAdSB black hole then becomes
\begin{equation}\label{en1}
S_\text{BH} = \frac{1}{4\kappa_\text{H}}\int_0^\pi \int_0^{2\pi}(\triangledown^0\xi^1)\frac{r_+^2}{h_+^5}\sin \theta d\theta d\phi = 
\pi x_+^2 =  \frac{\pi r_+^2 R^2}{(r_b^2-r_+^2)}.
\end{equation}
Note that the horizon area of the SAdSB black hole is calculated by the surface integration
\begin{eqnarray}\label{25}
\mathcal{A}_H = \int_0^{\pi} \int_0^{2\pi} \sqrt{g_{\theta \theta}g_{\phi \phi}}d\theta d\phi = 4\pi x_+^2 = 4\pi\frac{r_+^2 R^2}{r_b^2-r_+^2}.
\end{eqnarray}
By comparing \eqref{en1} and \eqref{25}, we remarkably realize that the entropy of SAdSB can be written in the usual form of the Bekenstein-Hawking area law as
\begin{equation}\label{en2}
S_\text{BH} = \frac{\mathcal{A}_H}{4}.
\end{equation}
%According to the first law of thermodynamics, we can obtain the Hawking temperature in the following
%\begin{eqnarray}
%T_\text{H}=\frac{\partial M}{\partial S_\text{BH}}   
 %= \frac{L^2+3x_+^2}{4\pi L^2x_+} \equiv \frac{\kappa_\text{H}}{2\pi}
%\end{eqnarray}
Combining \eqref{noe2},\eqref{en}, and \eqref{en2} and eliminating $S_\text{BH}$, we can write the Smarr formula in term of black hole mechanics as
\begin{eqnarray}\label{sma}
\frac{M}{2} = \frac{\kappa_\text{H}}{8\pi}\mathcal{A}_H-\frac{x_+^3}{2R^2}.
\end{eqnarray}

\section{Thermal Phase Transition of the Schwarzschild-Anti de Sitter-Beltrami Black Hole}

%review AdS Temperature bla bla .....

 In this section, we are interested to determine the thermodynamic quantities and thermodynamic relation of the SAdSB black hole. The phase structure and phase transition of this black hole will be studied by the obtained heat capacity and Gibbs free energy.
% From the non-inertial AdS metric \eqref{AdS coordinates} we use the coordinate transformation 
%\begin{equation}
%\frac{R}{L} = \sqrt{1+\frac{r_n^2}{L^2}}, 
%\end{equation}
%and Euclideanization of the time coordinate $t_n \rightarrow i\tau_n$, so we get the Euclidean non-inertial AdS metric as
%\begin{equation}
%ds^2_{E} = R^2\left(d(\frac{\tau_n}{L})\right)^2+\frac{L^2}{r_n^2}dR^2   
%\end{equation}
Let us formulate the thermodynamic formalism from the entropy area law \eqref{en2} and the mass relation \eqref{23a}. The curvature radius of the AdS spacetime $R$ can be represented as the thermodynamic pressure by the relation \cite{Kubiznak:2016qmn,Kubiznak:2012wp,Dolan:2012jh,Dolan:2014mra}
%write about enthalpy tempature volume
\begin{equation}\label{press}
P = \frac{3}{8\pi R^2}.
\end{equation}
From the entropy area law shown in \eqref{en1} and the pressure relation in \eqref{press} we propose that the mass of the SAdSB black hole \eqref{23a} can play a role as the enthalpy function $H(S_\text{BH},P)$ \cite{Kastor:2009wy,Kubiznak:2016qmn,Dolan:2012jh} expressed as
\begin{equation}
M = H(S_\text{BH},P) = \sqrt{\frac{S_\text{BH}}{4\pi}}\left(1+\frac{8S_\text{BH}P}{3}\right).
\end{equation}
Then we can use the Maxwell's thermodynamic relation to determine the Hawking temperature and the thermodynamic volume of the SAdSB black hole as 
\begin{equation}\label{hth}
 T_H = \frac{\partial H}{\partial S}\bigg|_{P} = \sqrt{\frac{1}{16 \pi S_\text{BH}}}(1+8PS_\text{BH}),
\end{equation}
\begin{equation}\label{vol}
V  = \frac{\partial H}{\partial P}\bigg|_{S_\text{BH}} = \frac{4}{3}\frac{S_\text{BH}^{3/2}}{\sqrt{\pi}}.  
\end{equation}
From \eqref{hth} we can show that the Hawking temperature is related to the surface gravity at the event horizon of the SAdSB black hole \eqref{kh} as
\begin{eqnarray}\label{1bhth}
T_\text{H}   
 = \frac{R^2+3x_+^2}{4\pi R^2x_+} = \frac{1}{4\pi r_+ R}\left(\frac{r_b^2+2r_+^2}{\sqrt{r_b^2-r_+^2}}\right) = \frac{\kappa_\text{H}}{2\pi},
\end{eqnarray}
and the thermodynamics volume \eqref{vol} is expressed through the event horizon radius $r_+$ in the formula
\begin{equation}
V = \frac{4\pi x_+^3}{3} = \frac{4}{3}\frac{\pi r_+^3 R^3}{(r_b^2-r_+^2)^{3/2}}.   
\end{equation}

\begin{figure*}[!ht]
	\begin{tabular}{c c}
		\includegraphics[scale=0.35]{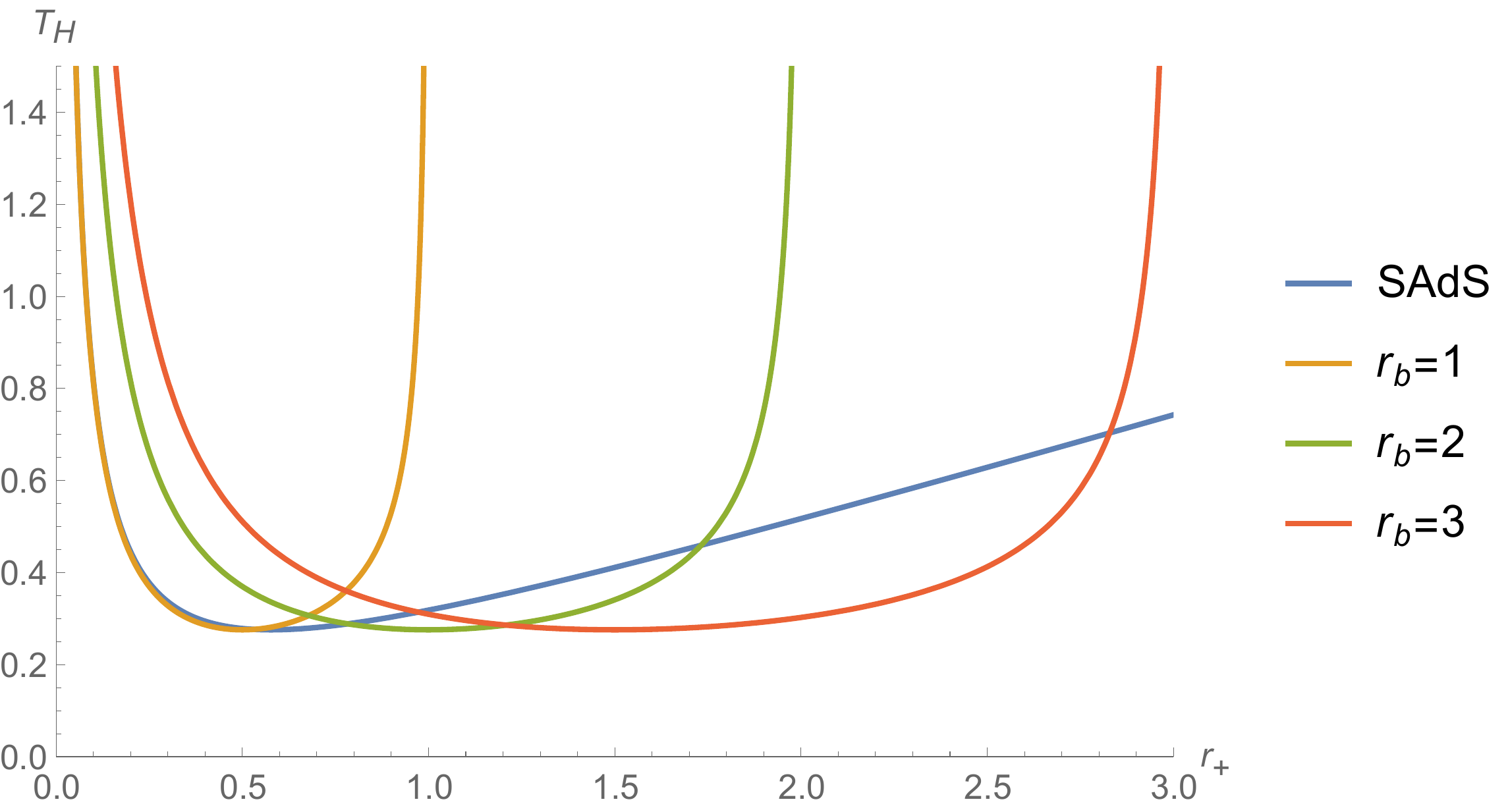}\quad
		\includegraphics[scale=0.35]{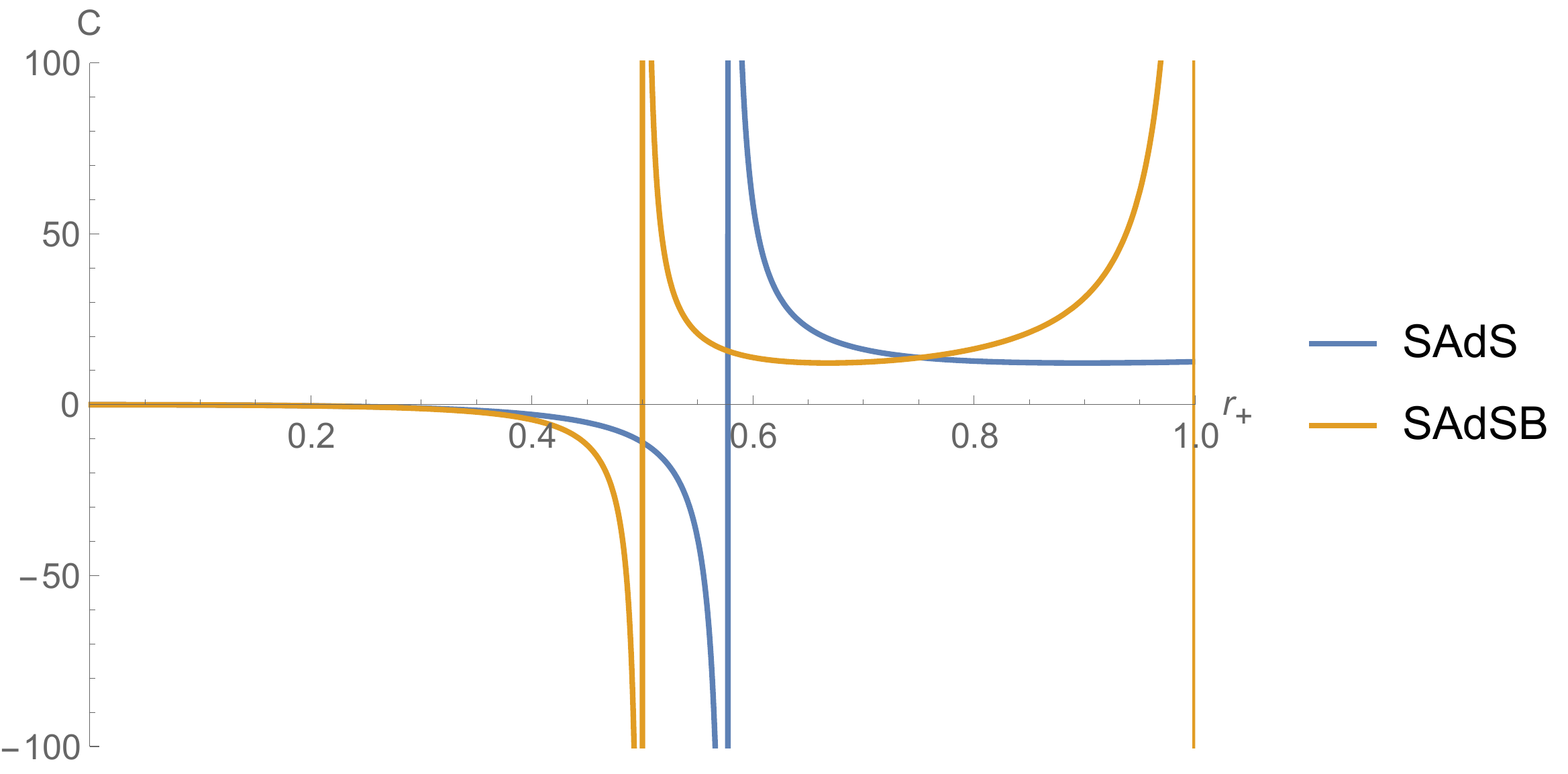}
	\end{tabular}
	\caption{Left: The graph between the Hawking temperature $T_H$ versus the event horizon radius $r_+$ for a fixed value of AdS radius $R=1$ is plotted with $r_b =1$ (orange), $r_b=2$ (green) and $r_b=3$ (red), compare with the case of SAdS black holes (blue). Right: Heat capacity of the SAdS black holes (blue) and SAdSB black holes with $r_b=1$ (orange).}\label{fig:SAdSB}
\end{figure*}

Therefore, from \eqref{sma}, we can form the first law of black hole mechanics describing the relation between the mass, event horizon area and pressure of the SAdSB black hole as  
\begin{eqnarray}\label{1bhm}
dM = \frac{\kappa_\text{H}}{8\pi}d\mathcal{A}_H+VdP.
\end{eqnarray}
We notice that the Smarr relation in \eqref{sma} and the first law are presented in the same way as in the case of the SAdS black hole \cite{Kubiznak:2016qmn}.
As a result, the first law of black hole thermodynamics in the Anti-de Sitter-Beltrami spacetime is formulated in the such formula
\begin{eqnarray}
dM = dH(S_\text{BH},P) = T_\text{H} dS_\text{BH}+VdP.
\end{eqnarray}
Now we can demonstrate the thermodynamic properties of the SAdSB black hole. The Hawking temperature (\ref{1bhth}) shows that the region of the SAdSB Black hole is defined only under the boundary region of $(0,r_b)$ as shown in Fig~\ref{fig:SAdSB}  by the relation $h^2 = 0$, because the temperature value is bounded with the limit of the radius $r_b$. This is different from the case of the SAdS black hole temperature but similar to the result of the Schwarzschild black hole in a box \cite{Basu:2016srp}. The SAdSB black hole temperature is slightly changed in the limit $R \gg t$.

To study of the phase transition of the SAdSB black hole, it is known that a phase of a black hole is characterized by heat capacity and Gibbs free energy.
The heat capacity of the SAdSB black hole is given by
\begin{equation}
C = \frac{\partial M}{\partial T_H}=\frac{2\pi r_+^2 R^2(r_b^2+2r_+^2)}{(r_b^2-r_+^2)(4r_+^2-r_b^2)}.
\end{equation}
From the graph between the temperature and the horizon radius, we see that the minimal temperature $T_{min}$
characterizes two states of the SAdSB black hole. Exactly, when we plot the graph between the heat capacity and the horizon radius, there is a discontinuity at the horizon radius $r_{c} = \displaystyle\frac{r_b}{2}$ which indicates the phase transition between the two phase of SAdSB black hole. %Length Scale of Phase Transition
In the limit $R \gg t$ the value of the $r_c$ approximately equal to $R/2$, which is less than the critical radius $r_c$ of the SAdS black hole. Similarly, the minimal temperature is calculated used the critical radius $r_c$ as
\begin{equation}
T_{min} = \frac{\sqrt{3}}{2\pi R}.
\end{equation}
This temperature characterizes the SAdSB black hole into two different states according to its heat capacity, i.e. small black holes for $C<0$ and large black holes for $C>0$. The minimal temperature of the SAdSB black hole is the same as the case of the SAdS black hole. The two solutions of the horizon radius for the small ($r_s$) and the large ($r_l$) black hole states are temperature dependent and can be expressed as
\begin{equation}
r^2_{l,s} = \frac{r_b^2\left(4\pi^2 R^2 T_H^2-1 \pm 2\pi R T_H\sqrt{4\pi^2 R^2 T_H^2-3}\right)}{2(1+4\pi^2 R^2 T_H^2)}.
\end{equation}

Moreover, the phase transition of the SAdSB Black hole can be described by the Gibbs free energy that is determined by 
\begin{equation}\label{gfe}
G = M-T_H S_{BH} = \frac{r_+ R(r_b^2-2r_+^2)}{4(r_b^2-r_+^2)^{3/2}}.
\end{equation}

\begin{figure}[h!]
 \begin{center}
  \includegraphics[scale=0.5]{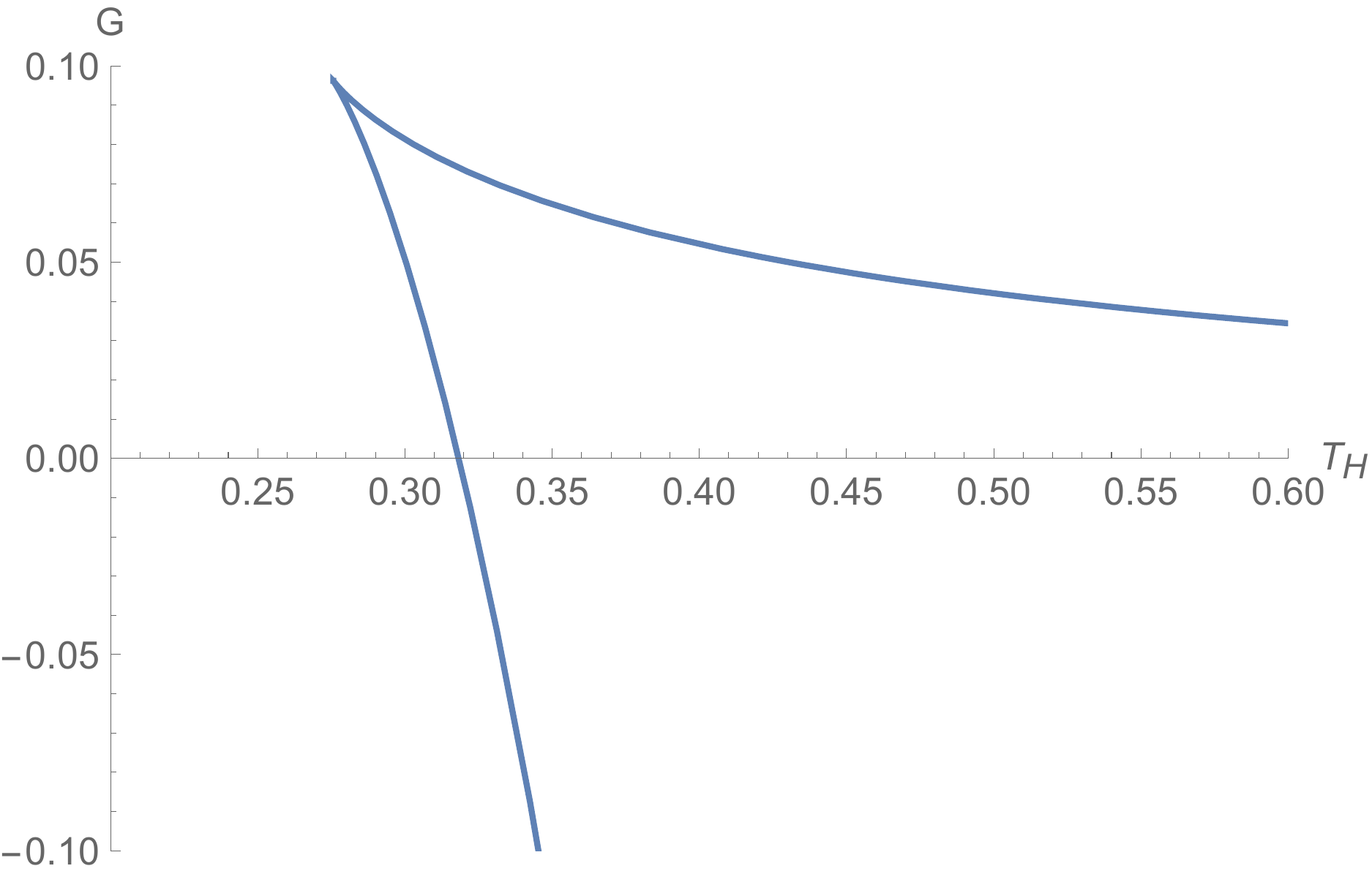}
   \end{center}
    \caption{Gibbs Free Energy of the SAdSB Black Hole}
    \label{fig:FTSAdSB}
\end{figure}
From the graph between the Gibbs free energy and the Hawking temperature in Fig~\ref{fig:FTSAdSB}, the discontinuity in the slope due to the phase transition of the SAdSB Black hole also emerges here. From the graph ones can obviously see that the discontinuity of the Gibbs free energy is located at the critical minimal temperature $T_{min}$, which agrees well with the result of the previous discussion. We, therefore, can also use the relation between Gibbs free energy and the Hawking temperature to imply the existence of the two phases of the SAdSB black hole. Furthermore, we consider the values of the Gibbs free energy of the large and small black hole limits.
The small black hole limit occurs under the condition $r_+,t \ll R$, with its Gibbs free energy of
\begin{equation}
G \sim 16\pi^3 R^4 T_H^3 r_+^3.
\end{equation}
The large black hole limit, on the other hand, appears when $r_+ \sim R \gg t$, and its Gibbs free energy becomes
\begin{equation}
G \sim -\frac{16\pi^3 R^4 T_H^3}{27}.
\end{equation}
Obviously, the large black hole limit in SAdSB is different from those in SAdS as the size of the large black hole is nearly the AdS radius.

Of course, from the Fig~\ref{fig:FTSAdSB} the Hawking-Page temperature of the SAdSB Black hole can be specified by the intercept $G=0$ of the graph between free energy and temperature of the SAdSB black hole
\begin{equation}
T_{HP} = \frac{1}{\pi R}.
\end{equation}

It can thus be interpreted that the first order phase transition, from thermal AdS spacetime to a large black hole in Beltrami coordinates, occurs at this temperature and at the radius $r_{HP} = \frac{r_b}{\sqrt{2}}$. Similarly, in the limit $R \gg t$ the value $r_{HP} \simeq R/\sqrt{2}$ for the SAdSB black hole is also less than the radius of the Hawking-Page transition of the SAdS black hole.

Let us consider the canonical ensemble of the SAdSB in the contexts of the off-shell Gibbs free energy \cite{PhysRevD.102.024085,PhysRevD.104.064004}. The off-shell Gibbs free energy can be obtained by replacing the Hawking temperature $T_H$ with the ensemble temperature $T$, which gives 
\begin{equation}
\bar{G}=M-TS = \frac{R r_+ r_b^2}{2(r_b-r_+)^{3/2}}-\frac{\pi T R^2 r_+^2}{r_b^2-r_+^2}.
\end{equation}

\begin{figure}[h!]
 \begin{center}
  \includegraphics[scale=0.5]{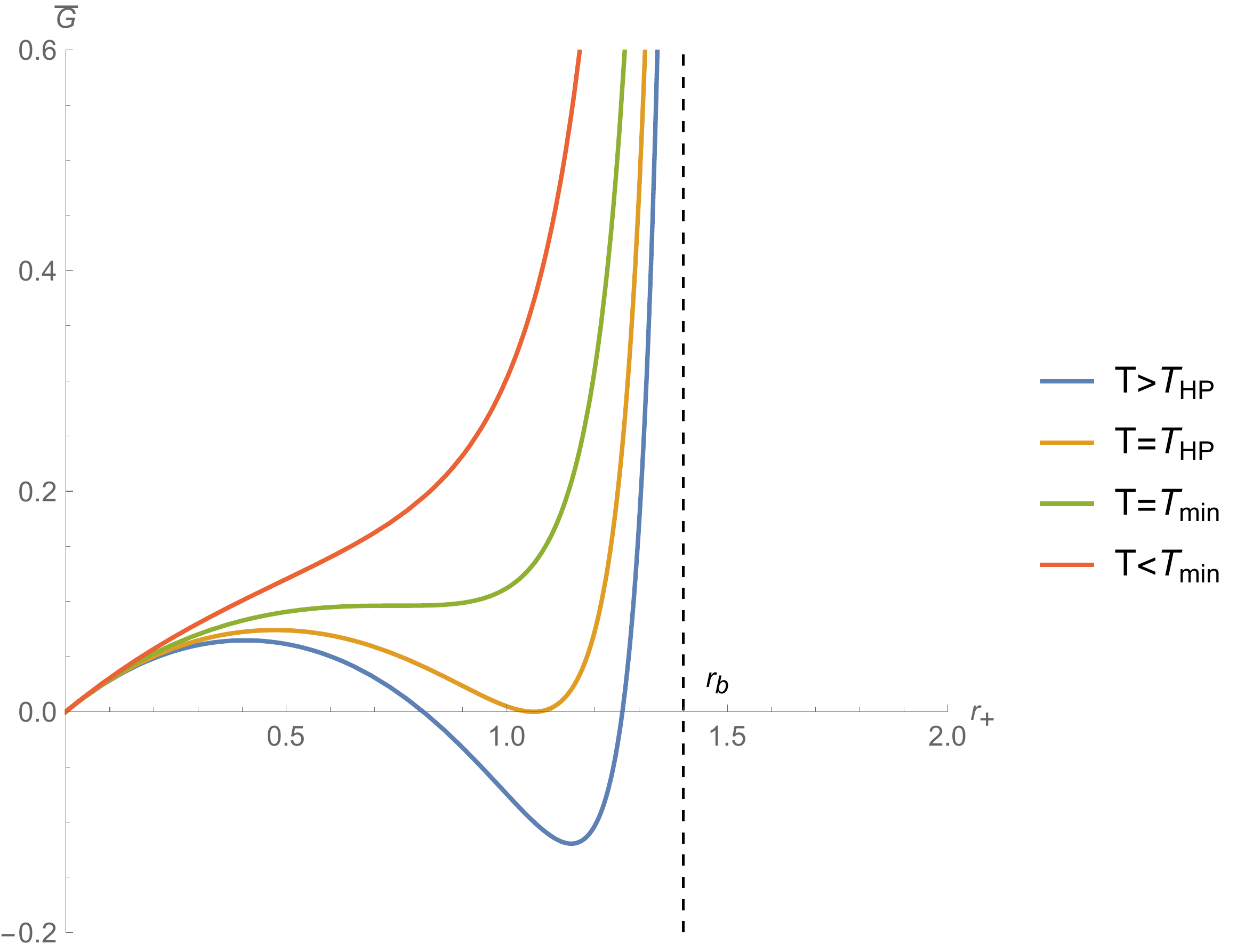}
   \end{center}
    \caption{Gibbs Free Energy of the SAdSB Black Hole with the bounded radius $r_b = \sqrt{2}$}
    \label{fig:FrhAdSB}
 \end{figure}
 
The graph of $\bar{G}$ as the function of $r_+$ is shown in the Fig~\ref{fig:FrhAdSB}. Notice that $\bar{G}$ of the SAdSB black hole is bounded within the region where $r_+<r_b$, in the same way as the case of the asymptotically flat Schwarzschild black hole in a cavity \cite{PhysRevD.102.024006}. At low temperature $T<T_{min}$, there is no black hole phases and the only one global minimum of $\bar{G}$ is the thermal radiation in an AdSB spacetime. When the temperature increases to $T=T_\text{min}$, the black hole configuration appears at the inflection point. For $T_\text{min}<T<T_\text{HP}$ two black hole configurations appear, namely small and large black hole phases, which correspond to the local maximum and minimum of $\bar{G}$, respectively. At the Hawking-Page phase transition temperature $T=T_\text{HP}$, the off-shell Gibbs free energies of both radiation phase and large black hole phase are equal, namely $\bar{G}_\text{rad}=\bar{G}_\text{LBH}=0$, as shown in the orange curve in the Fig~\ref{fig:FrhAdSB}. At this point, the order parameter $r_+$ discontinuously changes from the thermal radiation to large black hole phase, and therefore, this phase transition is regarded as the first order type of phase transition. Finally, in the $T>T_\text{HP}$ region, we have found that the large black hole phase is the most thermodynamically preferred state in the free energy landscape of the system. 
\section{Conclusion}
In this work, we have realized the solution of the gravitating mass in the inertial frame of the Anti-de Sitter spacetime. The event horizon of the SAdSB black hole is determined by a zero-norm of the Beltrami-time translation Killing vector in \eqref{kill1}. We find that the entropy of the SAdSB black hole can be determined from the surface area of the event horizon which satisfies the Bekenstein-Hawking area law. Remarkably, the entropy can be regarded as the Noether charge generated by the Killing vector.  The Smarr relation for this black hole also has been formulated by Noether charge integral, and the thermodynamic quantities have been determined from the given entropy.  

Apparently, the Hawking temperature and Gibbs free energy depend on the bounded radius and their character is analogous to a flat black hole in a cavity. However, the cavity boundary of the SAdSB black hole varies in time. This is different from the SAdS black hole as there exists the time-dependent boundary radius in the SAdSB black hole. Moreover, the critical radius $r_c$ for occurrence to phase transition of small-large black hole in the AdSB spacetime is less than the radius $r_c $ in the AdS spacetime, but in the both spacetime the minimal temperature $T_{min}$ takes the same value. 

We also discussed about the Hawking-Page phase transition between a thermal AdS background and large black hole state in Beltrami coordinates. The given phase-transition temperature $T_{HP}$ remains as the case of the SAdS black hole, whereas the transition radius $r_{HP}$ of the SAdSB black hole decreases which respect to the black hole in non-inertial coordinates. We obtained the off-shell free-energy landscape which reveals all possible phases of the SAdSB black holes. We expect similarities between phases of the SAdSB black hole and the phases of a black hole in a box or a cavity. 

We notice that changing the sign of the spacetime variables of SAdSB black hole induces the phase transition as indicated by the discontinuity in slope of the graph of Gibbs free energy and temperature. The phase transition, on the other hand, does not exist in the SdSB black hole \cite{Liu:2016urf}, as the relation between the Gibbs free energy and the temperature has a continuous slope.

\begin{acknowledgments}
 The authors would like to thank C. Promsiri, S. Ponglertsakul and S.N. Manida for helpful discussions of this article.
\end{acknowledgments}
 %discus phase transรtion of the SAdSB BH. It is interesting to explore the duality between SAdSB and gauge theories at the boundary in the AdS/CFT context 

\appendix

\section{Thermodynamics of the Schwarzschild Black Hole in Anti-de Sitter Spacetime}

\qquad In this appendix, we review the thermodynamics of the SAdS black holes. We use the method of the Noether charge to calculate the thermodynamics quantities for the SAdS black hole, and obtain the Smarr relation. It is known that the line element of the SAdS spacetime can be written as
\begin{eqnarray}\label{SAdS1}
 ds^2=f(r_n)dt_n^2-\frac{dr_n^2}{f(r_n)}-r_n^2d\Omega^2_2 \ , \ \ \ f(r) = 1+\frac{r_n^2}{R^2}-\frac{2 M}{r_n},
\end{eqnarray}
where $d\Omega^2_2 = d\theta^2+\sin^2\theta d\phi^2$ is the square of line element on a two-sphere. The parameters $M$ and $R$ are the mass of the black hole and the AdS radius, respectively.

The time translation Killing vector of the SAdS spacetime is represented as
\begin{equation}\label{killa}
 \xi_n  = \frac{\partial}{\partial t_n}.
\end{equation}
So we can rewrite this vector in the component form as
\begin{equation}
(\xi_n){}^\mu = (1,0,0,0).
\end{equation}

The SAdS black hole solution has an event horizon at $r_n = r_+$ such that $f(r_+) = 0$ with the condition of the zero-norm of the Killing vector \eqref{killa}. So the black hole mass $M$ relates to the event horizon radius as follow
\begin{eqnarray}
M = \frac{r_+}{2}\left(1+ \frac{r_+^2}{R^2}\right).
\end{eqnarray}

From the equation (\ref{Q}) we can find the non-zero component of the Noether charge element of the SAdS black hole which is expressed that
\begin{equation}\label{noe3}
Q^{01} = -Q^{10} = \frac{1}{16\pi}(\partial_1 g_{00}) =\frac{1}{8\pi}\left(\frac{r_n}{R^2}+\frac{M}{r_n^2}\right).
\end{equation}
Because of the conserved quantity is defined by the Noether charge integral over the event horizon region $\partial{\Sigma}$ in the formula \cite{PhysRevD.50.846, Frolov:1998wf}
\begin{equation}\label{noe4}
Q = \frac{1}{8\pi}\int_{\partial\Sigma}\nabla^{[\mu}\xi^{\nu]}d\sigma_{\mu\nu}
= \frac{1}{8\pi}\int_{\partial\Sigma}(\partial_1 g_{00})d\sigma_{01}.
\end{equation}
where $d\sigma_{01} = \frac{1}{2}r_n^2\sin\theta d\theta d\phi$ is the two-dimensional bifurcation surface of the SAdS black hole.

Thus, we put the equation (\ref{noe3}) to the integral (\ref{noe4}), the Noether charge of the SAdS spacetime at the event horizon radius
$r_n = r_+$ can be represented as
\begin{equation}\label{qsads}
Q_{SAdS} = \frac{1}{8\pi}\int_0^\pi \int_0^{2\pi}\left(\frac{r_+}{R^2}+\frac{M}{r_+^2}\right)r_+^2\sin\theta d\theta d\phi
= \frac{M}{2}+\frac{r_+^3}{2R^2}.
\end{equation}
It is known that the surface gravity at the event horizon of the SAdS black hole can be determined as
\begin{equation}\label{kappa2}
\kappa_H = \sqrt{-\frac{1}{2}\nabla^\nu \xi^\mu \nabla_\nu \xi_\mu} =  \frac{R^2+3r_+^2}{2R^2 r_+}.
\end{equation}
Thus we can show that the ratio of the Noether charge integral and the surface gravity is related to the area of the SAdS black hole's event horizon by
\begin{eqnarray}
\frac{2\pi Q}{\kappa_H} = \pi r_+^2 = \frac{\mathcal{A}_H}{4},
\end{eqnarray}
where
\begin{eqnarray}
\mathcal{A}_H = \int_0^{\pi} \int_0^{2\pi} \sqrt{g_{\theta \theta}g_{\phi \phi}} d\theta d\phi
= 4\pi r_+^2,
\end{eqnarray}
is the event horizon area of the SAdS black hole.

So we suggest that the SAdS black hole entropy also can be formulated from the Noether charge integral \eqref{qsads} which satisfies the Bekenstein-Hawking area law as
\begin{eqnarray}
S_{BH} = \frac{2\pi}{\kappa_H}Q_{SAdS}  = \pi r_+^2.
\end{eqnarray}

Since, for the SAdS black hole, the value $\frac{\kappa_H}{2\pi}$ belongs to the Hawking temperature expressed in the formula
\begin{eqnarray}
T_H = \frac{\kappa_H}{2\pi} = \frac{f'(r_+)}{4\pi} = \frac{R^2+3r_+^2}{4\pi R^2r_+}.
\end{eqnarray}
Therefore, the conserved quantity $Q_{SAdS}$ can be represented as the product of the surface gravity and event horizon area and it satisfies the relation \cite{PhysRevD.74.044007}
\begin{equation}
Q_{SAdS} = \frac{\kappa_H}{8\pi}\mathcal{A}_H = T_{H}S_{BH}.
\end{equation}
Furthermore, the Smarr relation can be formulated from the Noether charge, entropy and temperature as
\begin{equation}
\frac{M}{2} = \frac{\kappa_H \mathcal{A}_H}{8\pi}-VP = T_HS_{BH}-VP,
\end{equation}
where $P = \frac{3}{8\pi R^2}$ is the pressure term of the SAdS black hole and $V = \frac{4}{3}\pi r_+^3$ is the volume of the SAdS black hole \cite{Kubiznak:2016qmn,Kubiznak:2012wp,Dolan:2012jh}.
The first law thermodynamics has been constructed from the variation of the entropy and the Hawking temperature of the SAdS black hole in the equation
\begin{equation}
dM = T_HdS_{BH}+VdP.
\end{equation}
Remark that, this equation can also be obtained from the Komar integral as in \cite{Kubiznak:2016qmn}. 

Furthermore, we find the event horizon radius in term of the Hawking temperature as
\begin{eqnarray}
(r_+)_{l,s} = \frac{1}{3}\left( 2\pi R^2T_H \pm \sqrt{4\pi^2 R^4 T_H^2 - 3R^2} \right). \label{horizon}
\end{eqnarray}
There are two possible branches of black hole in AdS spacetime at a given temperature $T_H$, a large and small black holes correspond to positive ($(r_+)_l$) and negative ($(r_+)_s$) sign in \eqref{horizon}, respectively. These two branches of black holes are degenerated at the minimum temperature
\begin{eqnarray}
T_{min} = \frac{\sqrt{3}}{2\pi R},
\end{eqnarray}
and a horizon radius becomes $r_c = \frac{R}{\sqrt{3}}$. For $T<T_{min}$, the event horizon radius $r_+$ in \eqref{horizon} becomes a complex number, this mean that there are no black hole configuration below $T_c$ and the space is filled with thermal radiation.

\begin{figure}[h!]
 \begin{center}
  \includegraphics[scale=0.5]{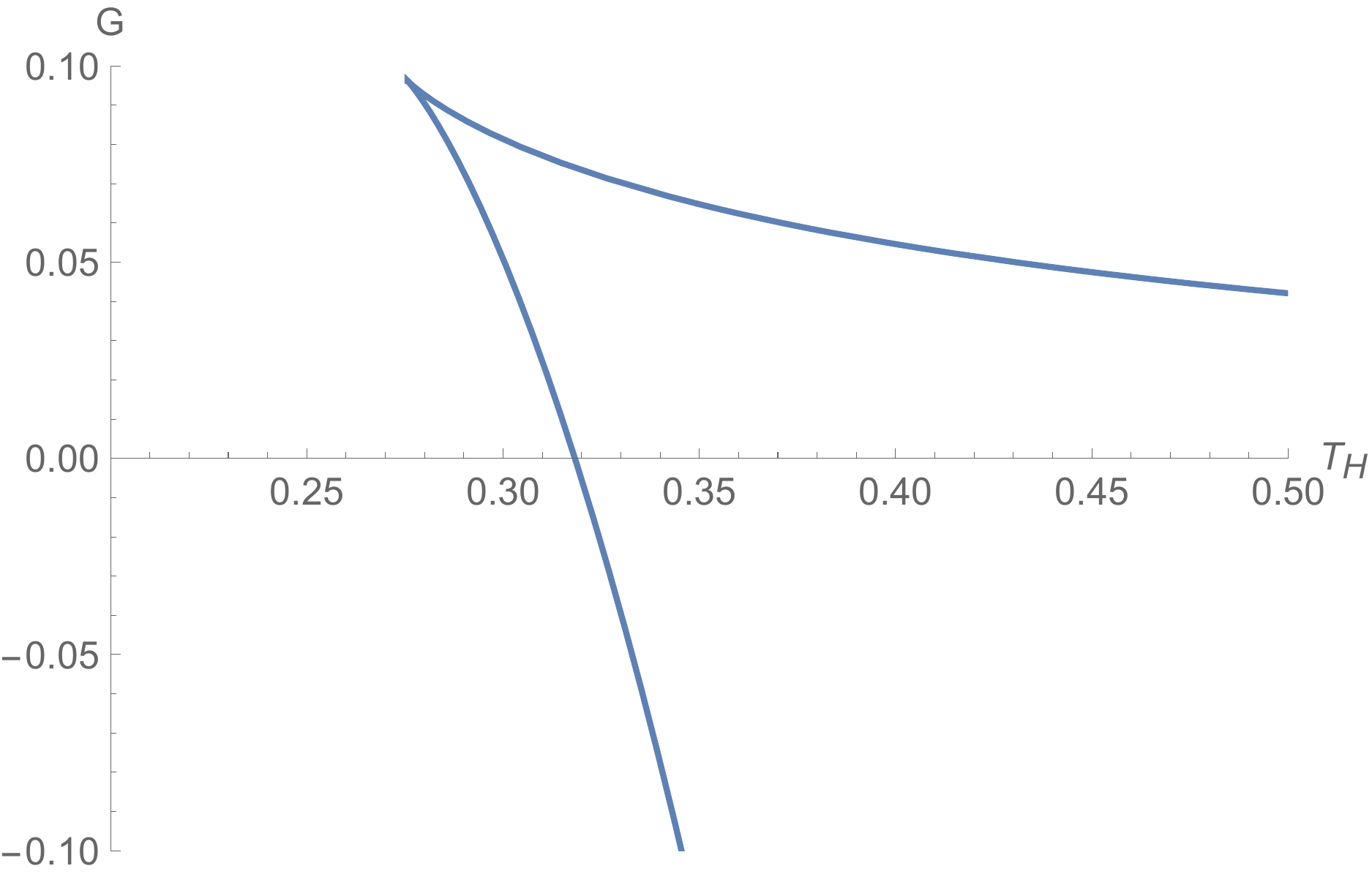}
   \end{center}
    \caption{Gibbs Free Energy of the SAdS Black Hole}
    \label{fig:FTAdS}
\end{figure}

The heat capacity of the SAdS black hole can be evaluated as the derivative of the mass by the Hawking temperature
as
\begin{eqnarray}
C = \frac{\partial M}{\partial T_H} = \frac{2\pi r_+^2(3r_+^2 + R^2)}{(3r_+^2 - R^2)}.
\end{eqnarray}
Obviously, the heat capacity is negative when $r_+<r_c$ and positive when $r_+>r_c$ which corresponds to small and large black hole respectively. To investigate the thermodynamic stability of three possible configurations, pure radiation, small and large black holes, we study the free energy as a function of the temperature. For SAdS black hole, the Gibbs free energy can be calculated from the formula $G = M - T_HS_{BH}$ as,
\begin{eqnarray}
G = \frac{r_+}{4}\left( 1 - \frac{r_+^2}{R^2} \right).
\end{eqnarray}
The graph of $G$ versus $T_H$ is shown in Fig~\ref{fig:FTAdS}.
We note that from the Fig~\ref{fig:FTAdS} the phase transition between small and large black holes is second-order because the second derivative of the free energy is discontinuous at the temperature $T = T_{min}$. Furthermore, the thermal radiation phase appears under the Gibbs free energy is zero. It is important that the Hawking-Page phase transition occurs when the Gibbs free energy changes from zero to negative values at $T = T_{HP} = 1/\pi R$ and the radius of the black hole is $r_+ = R$.
It is the first-order phase transition because there is the discontinuity in slope of the Gibbs free energy.

\bibliography{ref}
\end{document}